# Spin Polarization via Electron Tunneling through an Indirect-Gap Semiconductor Barrier

Subodha Mishra, Sunita Thulasi, and Sashi Satpathy
Department of Physics, University of Missouri, Columbia, MO 65211

We study the spin dependent tunneling of electrons through a zinc-blende semiconductor with the indirect X (or Δ) minimum serving as the tunneling barrier. The basic difference between tunneling through the Γ vs. the X barrier is the linear-k spin-orbit splitting of the two spin bands at the X point, as opposed to the $k^3$ Dresselhaus splitting at the Γ point. The linear coefficient of the spin splitting β at the X point is computed for several semiconductors using density-functional theory and the transport characteristics are calculated using the barrier tunneling model. We show that both the transmission coefficient as well as the spin polarization can be large, suggesting the potential application of these materials as spin filters.

PACS indices: 72.25.-b, 71.20.Nr, 73.63.-b

An important problem in spintronics is the development of spin polarized current sources. One of the candidates for this is the asymmetric nonmagnetic heterostructure based on the interface-induced Rashba spin-orbit coupling[1]: $H_{SO} = \alpha(\vec{\sigma} \times \vec{k})\hat{n}$, which results in a spin-dependent potential for the scattering of the electrons (here, $\vec{\sigma}$ is the Pauli spin operator, $\vec{k}$ is the electron momentum, $\hat{n}$ is normal to the interface, and α is the coupling strength). The spin-dependent potential in turn leads to a net spin polarization for the outgoing electrons tunneled through the asymmetric heterostructure.

However, Perel' and coworkers[2] have shown that the tunneling process is itself spin dependent and, even for a <u>symmetric</u> heterostructure, a non-zero spin polarization can occur. In their work, tunneling through a potential barrier originating from the Γ conduction minimum was considered and a large spin polarization was found for the incident electron energy below the Γ minimum. We extend the analysis to tunneling through a barrier formed by the indirect X or the Δ minimum. We find that both the transmission coefficient as well as the spin polarization can be large, suggesting the potential use of these materials as spin filters. The basic difference in the physics originates from the spin-orbit coupling term, viz., the $k^3$ Dresselhaus coupling[3] at the Γ point versus the linear-k coupling at the X point.

We consider tunneling through a barrier (Fig. 1), where the barrier material has the zinc blende structure with no inversion symmetry (e.g., AlAs). The basic origin of the spin polarization via tunneling lies in the spin dependent band structure of the barrier material. If the conduction bands are spin split, then an incident electron with energy in the gap will see a spin-dependent barrier, resulting in a spin-dependent tunneling probability.

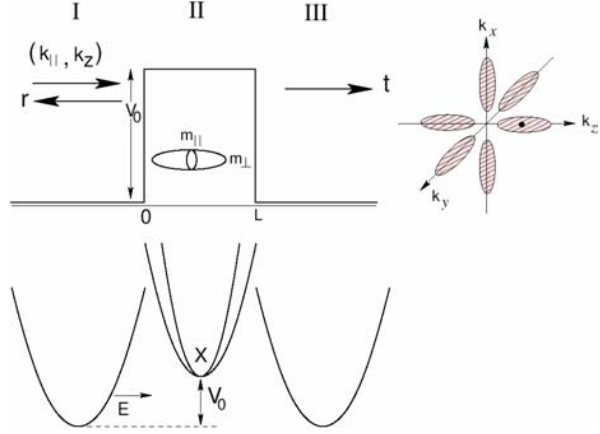

[Fig 1: Spin tunneling through a semiconductor barrier with an indirect minimum at the X point in the Brillouin zone.]

Consider the spin splitting of the electronic band structure. Quite generally, the time-reversal symmetry demands that the energy eigenvalues of an electron in a crystal must satisfy the condition:

$$E(\vec{k},\uparrow) = E(-\vec{k},\downarrow), \qquad (1)$$

where $\vec{k}$ is the Bloch momentum and $\uparrow,\downarrow$ denote the two spin states. If, in addition, the crystal has the inversion symmetry, then the eigenvalues remain unchanged if the Bloch momentum is inverted: $E(\vec{k},\uparrow) = E(-\vec{k},\uparrow)$. Thus when the inversion symmetry is present in the crystal, the energy eigenvalues become spin degenerate:

$$E(\vec{k},\uparrow) = E(\vec{k},\downarrow). \qquad (2)$$

This is the case for example in silicon. However, in crystals without the inversion symmetry, this is not the case, which leads to a spin splitting of the band structure.

In this work, we focus on materials such as AlAs in the zinc blende structure, where, in addition to the lack of the inversion symmetry, the conduction band minimum occurs at a point different from the center of the Brillouin zone, specifically, at the X point or at the Δ point along the Γ-X line.

The spin splitting at the X point for a zinc blende semiconductor is linear in k and is described by the Hamiltonian[3,4]

$$H_X = \beta(\sigma_x k_x - \sigma_y k_y) \qquad (3)$$

where the X point considered is along the $\hat{k}_z$ direction, X = 2π/a (001), which makes $(k_x, k_y)$ the deviation from the X point. Symmetry forbids spin splitting along the Γ-X axis, so that $k_z$ does not appear in the Hamiltonian. Here, 'a' is the lattice constant, $\vec{\sigma}$ is the Pauli spin operator, and β is a material dependent parameter. Diagonalization of Eq. (3) shows that the magnitude of the spin splitting is proportional to the magnitude of the parallel momentum $\vec{k}_{//}$:

$$E_\pm = \pm\beta\sqrt{k_x^2 + k_y^2} \equiv \pm\beta k_{\parallel}. \qquad (4)$$

The corresponding spin eigenfunctions $\chi_\pm(\sigma)$ depend on the direction of $\vec{k}_{//}$ and appear as the well-known spinwheels when plotted in the $(k_x, k_y)$ plane (Fig. 2). Expression (4) indicates that bands in the neighborhood of the X minimum are spin split, which results in a different barrier potential for the propagation of an electron with the two different spin eigenstates.

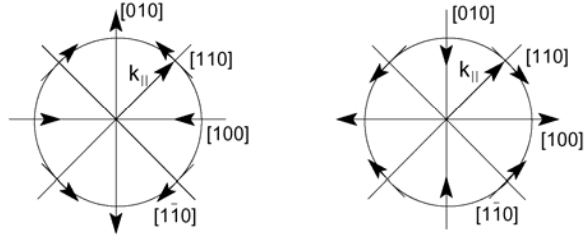

[Fig 2: Spin directions of the spin "up" and the spin "down" states corresponding to the energies $E_\pm = \pm\beta k_{//}$.]

The tunneling through a semiconductor is determined by its complex band structure. The momentum of the incoming electron can be split into an in-plane component $\vec{k}_{//}$, parallel to the interface, which is <u>real</u> and conserved during the entire tunneling process, and a component $k_z$ normal to the interface, which is not conserved and can be complex inside the barrier, $k_z = k_r + ik_i$, leading to the electron state decaying exponentially into the bulk as $\exp(-k_i z)$. We are concerned with the real energy solutions $E(\vec{k}_{//}, k_z)$ of the Schrödinger equation in the barrier region, with a real and fixed $\vec{k}_{//}$, and a complex $k_z$.

The analytic properties of the complex band structure have been analyzed by several authors, notably by Kohn and Heine,[5,6] and the topology of the complex bands has been studied numerically by Chang[7] for a number of semiconductors using a tight-binding model. We have illustrated in Fig. 3 the complex band structure for AlAs following the tight-binding results of Chang.

The complex bands emerge from the extrema of the band structure E(k) in real k space. In fact, the extrema are saddle points when E is plotted as a function of the complex momentum $(k_r, k_i)$, so that a maximum along $k_r$ is a minimum along $k_i$ and vice versa. In the vicinity of the saddle point at $k_0$, the energy E ($k_z$) has the form

$$E(k_z) \approx E_0 + (\hbar^2/2m_\perp)(k_z - k_0)^2, \qquad (5)$$

so that $E(k_z)$ is real along the Re k axis and on the line that cuts the Re k axis perpendicularly at $k_0$. Thus, for $\vec{k}_{//} = 0$, for instance, the momentum of the complex band attached to the Γ point is purely imaginary: $k_z = \pm ik_i$, while for the complex band attached to X, it is $X \pm ik_i$.

For the electron energies just below the X minimum, which is the energy region of interest in this paper, the important states for tunneling have the momentum $\vec{k} = (\vec{k}_{//}, k_z)$, with $k_z = X \pm ik_i$ and X = (2π/a) (0, 0, 1). These are the important tunneling states because they have the lowest values of the imaginary momentum and have therefore the slowest decay. As the electron energy is decreased further below the X minimum, the complex band associated with the Γ minimum takes over as the dominant state for tunneling, having now the slowest decay of all the complex states. The effect of the complex bands with larger imaginary momentum may be neglected as the wave functions for these states decay very rapidly, being negligible beyond just a few atomic layers from the interface. We are working in a regime where the X-derived states are the dominant ones and this logic has been used in writing down the envelope function part Eq. (8) of the wave function. Also, the parabolic approximation for the complex band structure Eq. (5) has been made in determining the normal component of the momentum in Eqs. (10) and (11).

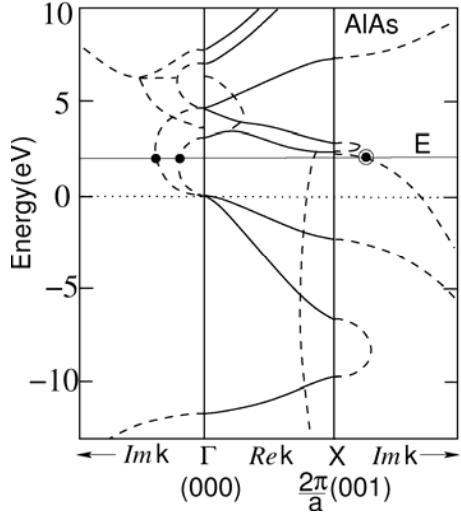

[FIG 3: Complex band structure $E(\vec{k}_{//},k_z)$ for AlAs with $\vec{k}_{//} = 0$. For energies in the gap, there are several tunneling states in the complex band structure (shown by heavy dots). The important states are those associated with the Γ point and the X valleys along the (001) direction. The complex bands associated with the L valleys as well as the other four X valleys located along $\hat{k}_x$ or $\hat{k}_y$ are not relevant for tunneling because the in-plane momentum $\vec{k}_{//}$ of the incoming electron is too small for these valleys to be accessible to the electron. For an energy just below the conduction minimum, the complex state denoted by the circled point is the dominant state for tunneling. For AlAs, the conduction minimum is at Δ, so that the momentum of the complex band attached to this point is given by $k_z = \mathrm{Re}\,k \pm i\mathrm{Im}\,k$, with $\mathrm{Re}\,k \approx \Delta$. The real part $\mathrm{Re}\,k$ is shown as a dashed line in the middle panel emerging from the conduction bottom.]

There is one last point that we need to consider regarding the wave function associated with the complex bands in the gap region. For this purpose, we consider a one-dimensional crystalline lattice with lattice constant 'a' in the nearly-free-electron approximation, taking $V_G$ as the Fourier component of the potential, where $G = 2\pi/a$ is the reciprocal lattice vector. This is a well-known approximation valid for a weak periodic potential, which allows one to obtain analytically the energies and the wave functions for the Bloch electrons. Its main virtue is in the illustration of the formation of the band gap in the solid.[8] The potential $V_G$ produces a gap at the Brillouin zone boundary at $X = \pm \pi/a$, spanning the region $\hbar^2 X^2/2m - |V_G| < E < \hbar^2 X^2/2m + |V_G|$. In this region, the electronic state has a complex momentum $k = \pi/a \pm iq$, with the imaginary part of the momentum given by $q = [\varepsilon V_G/((\hbar^2\pi^2/ma^2)+V_G)]^{1/2}$, where ε is the electron energy with respect to the conduction minimum at X. The electronic wave function just below the conduction minimum at X can be shown to be of the form

$$\Psi_k(z) \sim \cos(\pi z/a)\exp(\pm qz), \qquad (6)$$

which consists of an exponential part modulated by a cell periodic part that changes sign from cell to cell.

With these considerations, the wave function of the electron describing the tunneling process via the X minimum can be written as

$$\Psi_\pm(\vec{r},\sigma) = \chi_\pm(\sigma) e^{i\vec{k}_{//}\cdot\vec{r}_{//}} \phi_\pm(z) f_\pm(\vec{r}) \qquad (7)$$

where the ± refers to the two spin states, $\chi_\pm(\sigma)$ is the spin part, $e^{i\vec{k}_{//}\cdot\vec{r}_{//}} \phi_\pm(z)$ is the envelope function, and $f_\pm(\vec{r})$ is a cell periodic function. The latter is one on either side of the barrier, since we have vacuum there, and within the barrier material, it acquires a negative sign in going from cell to cell along the z direction. In general, if the complex bands originate from the Δ point (as shown in Fig. 3 for AlAs), the cell periodic part will acquire a complex phase factor, instead of a minus sign for the X point. The envelope function $\phi_\pm(z)$ in the three different region may be written as

$$\phi_\pm^I(z) = e^{ik_z z} + r_\pm e^{-ik_z z}$$
$$\phi_\pm^{II}(z) = (A_\pm e^{q_\pm z} + B_\pm e^{-q_\pm z}), \qquad (8)$$
$$\phi_\pm^{III}(z) = t_\pm e^{ik_z z}$$

where $r_\pm$ ($t_\pm$) is the reflection (transmission) coefficient. In terms of the effective masses and the energy of the X minima $V_0$ (see Fig. 1), the propagation vectors are given by

$$\frac{\hbar^2 k_z^2}{2m_\perp^I} = E - \frac{\hbar^2 k_{//}^2}{2m_{//}^I}, \qquad (9)$$

$$-\frac{\hbar^2 q_\pm^2}{2m_\perp^{II}} = E - V_\pm^{eff}, \qquad (10)$$

where the effective barrier height for tunneling $V_\pm^{eff}$ is given by the expression:

$$V_\pm^{eff} = (V_0 \pm \beta k_{//}) + \frac{\hbar^2 k_{//}^2}{2m_{//}^{II}}. \qquad (11)$$

In the envelope function approximation, the effect of the cell periodic part of the wave function is neglected. The reflection and transmission coefficients are then determined by matching the boundary conditions for the envelope functions, which amounts to satisfying the continuities of $\phi(z)$ and $(1/m_\perp)\, d\phi(z)/dz$ at the interfaces at $z=0$ and $z=L$. This leads to algebraic equations for the unknowns $A_\pm$, $B_\pm$, $r_\pm$, and $t_\pm$ appearing in Eq. (8), which can be solved to yield

$$T_\pm^{-1} = 1 + \frac{(E(m_\perp^{II}/m_\perp^I - 1) + V_\pm^{eff})^2}{4E(V_\pm^{eff} - E) m_\perp^{II}/m_\perp^I} \sinh^2 q_\pm L,$$

$$(12)$$

where $T_\pm \equiv |t_\pm|^2$ is the transmission coefficient. This expression is valid for the electron energy below the effective barrier potential $E < V_\pm^{eff}$; above it, the wave vector $q_\pm$ becomes imaginary per Eq. (10) and the hyperbolic sine function turns into a trigonometric sine function in Eq. (12).

In order to compute the transmission coefficients, the material parameters such as the effective masses and the spin-splitting parameter β at the conduction minimum are needed. While the experimental values for the effective masses are known, there exists, to our knowledge, no such results for β. We have therefore calculated the spin-splitting parameter β using the local density approximation (LDA) to the density functional theory and solving the Kohn-Sham density-functional equations using the linear muffin-tin orbitals (LMTO) method.[9] Earlier calculations by Christensen and coworkers[10] for the spin-splitting parameters at the Γ point using similar theoretical methods yielded results in reasonable agreement with experiments.

The starting point for these calculations is the relativistic Pauli equation, which is a two-component equation obtained from the four-component Dirac equation by eliminating the two small components. The Pauli Hamiltonian, which is correct to the second order in the fine structure constant α, adds the extra spin-orbit potential to the Kohn-Sham potential

$$H_{SO} = \xi(r) L.S, \quad (13)$$

where, for an electron moving in a central potential V(r), we have $\xi(r) = (\alpha^2/2r) dV(r)/dr$. It is often sufficient to treat this term using the first-order perturbation theory.

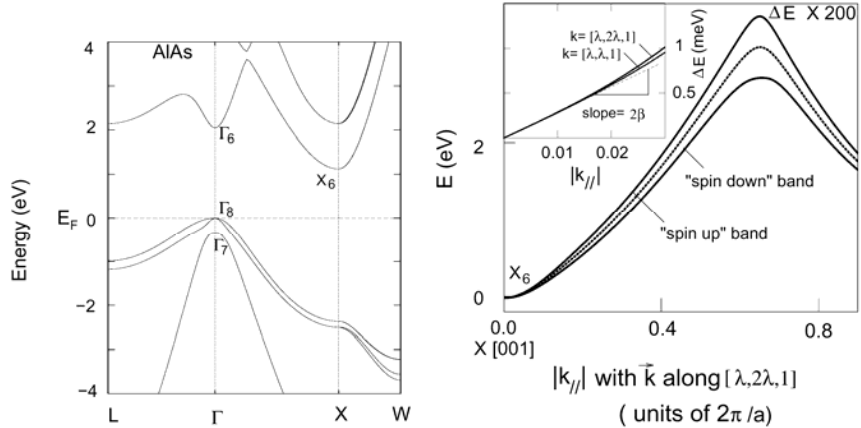

[Fig. 4. Density-functional band structure of AlAs (left) and the calculated spin-splitting for the lowest conduction band near the indirect minimum at X (right). The spin-splitting ΔE has been exaggerated by a factor of 200. The dashed line shows the spin-degenerate bands without the spin-orbit coupling term included. Inset shows the linear-k dependence of the spin splitting ΔE, which, in agreement with Eq. (4), is proportional to the magnitude of $\vec{k}_{//}$ only, irrespective of its individual components.]

The results from the density-functional calculations for the spin-splitting of the conduction minimum are shown in Fig. 4 for AlAs. The linear-k dependence of the spin splitting $\Delta E$ is verified from our results (displayed in inset of Fig. 4), with the calculated magnitude of the spin-splitting parameter $\beta \approx 10.9$ meV-Å. Note from the inset that the spin-splitting is proportional to the magnitude of the parallel momentum only and the proportionality constant $\beta$ is independent of the direction of the momentum line in the Brillouin zone in agreement with Eq. (2). Results for several other zinc blende semiconductors are shown in Table I. We have performed these calculations both without (standard LDA with the underestimated band gaps) as well as with the band-gap corrections (à la Christensen[11]) and have found only small differences in the computed spin splitting (e.g., 9.6 and 10.9 meV-Å for AlAs, with and without corrections, respectively). The standard LDA results without any band gap corrections have been presented in Table I and used throughout the paper.

Table I. Band structure parameters for some indirect-gap zinc-blende semiconductors. The spin splitting parameter β was obtained from density-functional calculations, while the rest of the parameters are experimental values taken from Ref. (12). The locations of the conduction minima are also according to Ref (12).

| material | AlP | AlSb | AlAs | GaP | Si |
|---|---|---|---|---|---|
| $\beta$ (meV-Å) | 10.3 | 6.2 | 10.9 | 72.4 | 0 |
| $m_\perp / m$ | 0.212 | 0.26 | 0.19 | 0.21 | 0.19 |
| $m_{//} / m$ | 3.67 | 1.0 | 1.1 | 7.25 | 0.92 |
| Indirect Gap (eV) ($\Gamma_v - X_c$) | 2.5 | 1.69 | 2.23 | 2.35 | 1.17 |
| Direct Gap (eV) ($\Gamma_v - \Gamma_c$) | 3.63 | 2.38 | 3.13 | 2.90 | 4.19 |
| Conduction minimum | X | $\Delta$ | $\Delta$ | $\Delta$ | $\Delta$ |

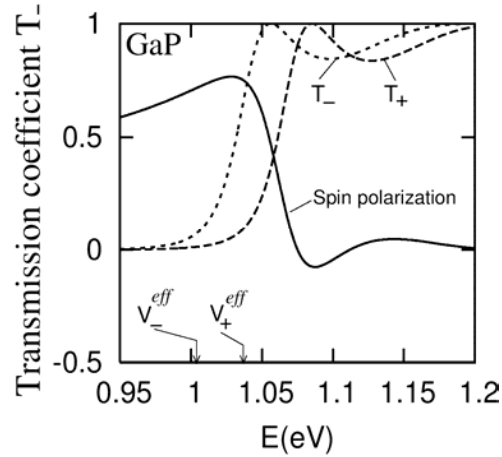

[Fig. 5. Transmission coefficients and the spin polarization as a function of the energy of the incident electron for GaP, as obtained from Eqs.
(12) and (14). The parameters used are: $V_0 = 1$ eV, $L = 60$ Å, and $k_{//} = 0.2$ Å$^{-1}$.]

With the parameters listed in Table I, we have computed the transmission coefficients for the two spin states, which are presented in Fig. (5). The maximum transmission for the two spins occur at slightly different energies resulting in a large spin polarization

$$P = \frac{T_- - T_+}{T_- + T_+}, \qquad (14)$$

shown by a solid line in Fig. (5). In a region spanning several meV's below the conduction minimum, only one spin type is transmitted, while the other spin type is mostly reflected back, leading to a robust value for the spin polarization. Fig. 6 shows the transmission coefficient and the net spin polarization for a number of indirect-gap semiconductors. The maximum polarization occurs for GaP, which originates from the large spin-splitting parameters β as seen from Table I.

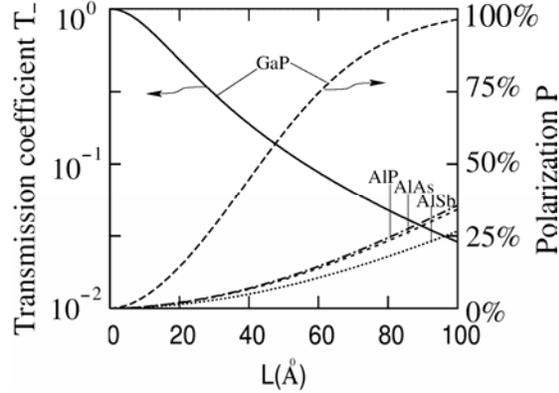

[Fig. 6. Transmission coefficient for the transmitted majority spins in GaP (solid line) and the corresponding spin polarization (dashed lines) for a number of zinc blende semiconductors as a function of the barrier thickness L. The energy of the incident electron is taken to be 1 meV below the effective barrier potential $V_-^{eff}$ for the lower spin state.]

We note that for many zinc blende semiconductors, the conduction minimum does not occur exactly at the X point, but close to it at the Δ point, along the Γ-X line. The effect discussed here does not change if the minimum occurs at Δ rather than at X, since the spin splitting turns out to be still robust along the entire Γ-X line, except of course at the Γ point, where symmetry forbids the linear-k term, causing the Dresselhaus $k^3$ term to take over. We have in fact computed the spin splitting along the entire Γ-X line using density-functional methods and have found the splitting to vary linearly with $k_{//}$ along the entire line. But because of the change of the orbital character of the wave function as one moves from Γ to X, the linear spin-splitting coefficient $\beta(\Delta) \equiv lim_{k_{//} \to 0} \delta E(k_{//}, k_z = \Delta)/\delta k_{//}$ turns out to depend very strongly on the k point on this line. The linear k dependence along the entire Γ-X line is consistent with the symmetry analysis in the early work of Dresselhaus.[3]

In conclusion, we have pointed out that the process of tunneling through an indirect conduction minimum of a semiconductor barrier can result in both a strong transmission coefficient as well as a strong spin polarization. It is not necessary to have the conduction minimum at the X point serving as the tunnel barrier and a minimum at Δ, often the case for the zinc blende semiconductors, is sufficient, since there is a strong spin splitting along the entire Γ-X line. We find that a barrier made up of GaP may be especially suitable to observe this effect on account of the strong spin-orbit coupling present there.

We thank Z. Popovic for help with the computations and Y. C. Chang for providing us with the tight-binding complex band structures for several semiconductors. We acknowledge support of this work by the U. S. Department of Energy through Grant No. DE-FG02-00ER45818.